# MiBoard: A Digital Game from a Physical World

Kyle B Dempsey, G. Tanner Jackson, Justin F. Brunelle, Michael Rowe, Danielle S. McNamara

Department of Psychology
Institute for Intelligent Systems
University of Memphis
Memphis. TN 38152
{kyle.b.dempsey, gtannerjackson, godorito, dsmcnamara1} @ gmail.edu, jbrunelle@cs.odu.edu)

## Abstract

Increasing user engagement is constant challenge for Intelligent Tutoring Systems researchers. A current trend in the ITS field is to increase engagement of proven learning systems by integrating them within games, or adding in game like components. Incorporating proven learning methods within a game based environment is expected to add to the overall experience without detracting from the original goals, however, the current study demonstrates two important issues with regard to ITS design. First, effective designs from the physical world do not always translate into the digital world. Second, games do not necessarily improve engagement, and in some cases, they may have the opposite effect. The current study discusses the development and a brief assessment of MiBoard a multiplayer collaborative online board game designed to closely emulate a previously developed physical board game, iSTART: The Board Game.

## Introduction

Games and game-based environments constitute an area of rapid growth in private, public, and research sectors. In 2007, while industries such as music and movies saw either negative or stagnant growth (-10.0% and +1.8% respectively), the gaming industry reported dramatic growth (+28.4%; Combs, 2008). Researchers of Intelligent Tutoring Systems (ITSs) have begun to leverage the engagement and appeal of games by incorporating game-like features within learning environments (McNamara, Jackson, & Graesser, 2009).

While it is intuitively clear that games are engaging and can often sustain interest over extended periods of time, it is still relatively unclear how this process occurs and which specific features are essential to the essence of games. Previous research has attempted to identify and investigate specific gaming components such as challenge, fantasy, complexity, control, rules, strategy, goals, competition, cooperation, and chance (Crookall, Oxford, & Saunders, 1987; Garris, Ahlers, & Driskell, 2002; Malone, 1981). However, these components have been primarily observed within the context of entertainment games. Only recently have these components been implemented and observed (and sometimes tested) in the context of learning environments (Barab et al., 2005; Johnson et al., 2003). Establishing the effects of game components on learning and motivation is important for those who are interested in developing systems that maximize learning benefits in computer-based systems. ITS developers and researchers often struggle to create just the right balance between implementing effective learning practices, while at the same time enhancing motivational aspects of the learning environment (Boyer et al., 2008; Jackson & Graesser, 2007). The principal goal of ITS technologies is most often to produce significant learning gains (e.g., learn a new skill or understand concepts within a specific domain). However, these systems, though often effective at producing learning gains, are sometimes uninspiring to those who use them. Focusing on maximizing learning benefits can suffice for experimental purposes, but it creates a problem for systems that are used repetitively and over long periods of time. Additionally, improving motivational aspects of learning environments is likely to produce indirect gains in learning, particularly if the modifications result in heightened engagement on the part of the learner (Graesser, Hu, & McNamara, 2005).

The intersection of these two fields (games and ITSs) provides a fertile ground to develop effective learning environments that maximize learning while at the same time fully engaging the user and instilling a desire to interact with the system. The remainder of this paper describes a work-in-progress to develop a learning system that borrows effective design elements from both games and ITS technologies.

## iSTART

Interactive Strategy Training for Active Reading and Thinking (iSTART) is a web-based tutoring system designed to improve students' reading comprehension by teaching self-explanation strategies. The iSTART system was originally modeled after a human-based intervention called Self-Explanation Reading Training, or SERT (McNamara, 2004; McNamara & Scott, 2001; O'Reilly, Best, & McNamara, 2004). The automated iSTART system has consistently produced gains equivalent to the human-based SERT program (Magliano et al., 2005; O'Reilly, Sinclair, & McNamara, 2004; O'Reilly, Best, & McNamara, 2004). Unlike SERT, iSTART is web-based, and can potentially provide training to any school or individual with internet access. Furthermore, because it is automated, it can work with students on an individual level and provide self-paced instruction. iSTART also maintains a record of student performance and can use this information to adapt its feedback and instruction for each student. Lastly, the iSTART system combines pedagogical agents and automated linguistic analysis to engage the student in an interactive dialog and create an active learning environment (e.g., Bransford, Brown, & Cocking, 2000; Graesser, Hu, & Person, 2001; Graesser, Hu, & McNamara, 2009 Louwerse, Graesser, & Olney, 2002).

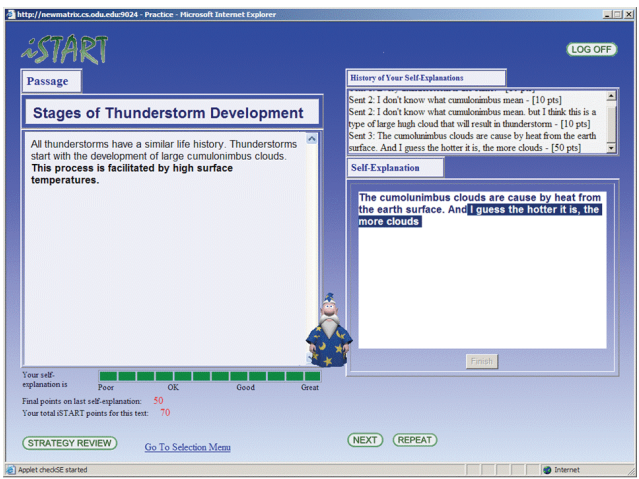

Figure 1: iSTART Interface

The key to success for iSTART is the integration of an extended practice module. The extended practice module in iSTART allows students to work with the system over a long-term interaction (over the course of a semester) and receive adaptive feedback for each self-explanation that they produce. This interaction requires time and practice, but fosters the development of deep knowledge. The mastery of content and learning strategies that will generalize to multiple contexts and tasks does not happen in hours, but rather in weeks, months, or years. Proficiency in content and strategies requires multiple sessions, across months of time. However, over time, this extended practice can become boring and tedious to students, particularly for those who most need such tutoring. Hence, a new alternative has been developed that incorporates game-based principles and offers a means of enhancing iSTART so it can be more appealing to students and engage them more frequently and seriously.

## iSTART: The Board Game

iSTART: The Board Game is an interactive multiplayer board game that utilizes the reading strategies espoused within iSTART. iSTART: The Board Game has two main game elements: competition between multiple players and movement around a physical board. These game elements differentiate iSTART: The Board Game from typical practice in three ways. First, game practice involves either three or four participants interacting with the same text and competing to be the player farthest along the path. These participants have a goal beyond just finishing the text (board progress). Second, the participants compete against one another to score points by self-explaining the text. Self-explanations are not only a means of progressing through a text, but also to advance one's rank in comparison to other players. In iSTART: The Board Game, the successful completion of a self-explanation provides the player the chance to roll the dice to advance around the board, score points, and select an event card.

The gameplay centers on a playing board, which is used to account for progress and compare performance with other players. Players attempt to move around the board using tokens as markers by rolling a pair of dice before a group-orienting monster finishes the game first (resulting in no winner). A single player's turn consists of 1) getting an assigned strategy, 2) self-explaining a sentence using that strategy, 3) other players guessing the strategy used, 4) revealing the assigned strategy, 5) discussion of strategy choices (if needed), 6) movement and tallying points (if correct), and 7) drawing an event card (if correct).

Players select a strategy card (with 2 optional strategies and respective pre-designated point values) and are expected to produce a self-explanation using either of the assigned strategies. After a player produces the self-explanation, the other players simultaneously attempt to identify the strategy used in the self-explanation (by secretly selecting a strategy identification card). Then all players show their strategy identification cards (and the reader reveals the assigned strategy card). If there is any disagreement, players discuss the strategies that they believe were used, and then revote on the strategy(ies). After completing the voting, players are awarded points based upon the agreement on the self-explanation strategy(ies) used. Players then use the points as a carry-over aspect of the board game (to use acquired items). In this board game, players roll the dice, move their piece, and then draw cards. The cards provide anything from additional movement instructions to power card abilities (used to affect either their own movements or other players' movements). Once the players finish with the board game portion of the round, the reader role switches to the next player, and the gameplay begins the next round.

Rowe (2008) conducted an assessment of iSTART: The Board Game. During the evaluation, 30 participants received training through an abbreviated version of iSTART (lecture portions only), played iSTART: The Board Game, and answered questions about their experience with the game. The questions were presented on a scale from 1 (Completely Disagree) to 6 (Completely Agree). Results in Table 1 show that participants found the game to be fun, useful, and easy to use.

Table 1. iSTART: The Board Game Evaluation Responses (ratings from 1 to 6, higher scores indicate agreement)

| Statement | Mean | SD |
|---|---|---|
| The Game Was Fun | 5.51 | 0.781 |
| The Game Improved Strategy Knowledge | 5.71 | 0.519 |
| The Game Was Hard To Use | 1.54 | 0.852 |

Unfortunately, iSTART: The Board Game is a physical presence board game and a separate entity from the computerized iSTART. Because of this difference in presentation media it is problematic to package the two together. Therefore, if a game such as this were to be included in a paired learning session, it would need to be computerized.

## MiBoard Game

MiBoard (Multiplayer interactive Board) Game is an online collaborative game directly adapted from iSTART: The Board Game. MiBoard was designed to serve as a potential alternative to iSTART's extended practice module. The central hypothesis is that game-based practice will be more engaging to the student (supported in Table 1), and possibly more effective in promoting learning, by incorporating motivational components that sustain concentration for long periods of time (Graesser, Chipman, Leeming, & Biedenbach, 2009; Moreno & Mayer, 2005).

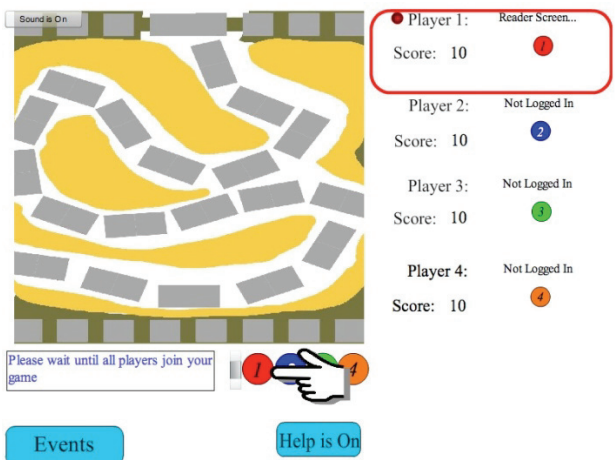

Figure 2: MiBoard Interface

MiBoard requires players to produce self-explanations as well as identify the strategies used in other players' self-explanations (just as in iSTART: the Board Game). The strategies included within MiBoard are those taught within the iSTART training (i.e., Comprehension Monitoring, Paraphrasing, Prediction, Elaboration, and Bridging). In MiBoard, users earn points when a majority of players agree upon the strategy used within a player's self-explanation (same as in iSTART: The Board Game). Players can spend these points during the game to change task parameters (e.g., change the assigned strategy) or activate special "in game" features (e.g., take an extra turn, freeze another player, draw an extra card, etc.). MiBoard does not provide feedback on the players' self-explanations. Instead, players receive feedback from the other players in the game through modeling of self-explanations as well as through a chat room discussion.

**Differences between the games**

While MiBoard is intended to directly mirror iSTART: The Board Game, there are noticeable differences between the two. Most importantly, MiBoard is computer-based. Players are required to use a computer interface to complete the game interactions, discussions, and voting. While this difference may seem innocuous, the challenges are great. For example, there is a wide range in computer skills among our target demographic ($9^{th}$-$12^{th}$ graders) with some users being relatively inexperienced with computers. These experience differences are likely to produce a wide variation in performance and pace (which causes the slowest user to dictate the speed of gameplay).

One of the game features excluded from MiBoard is the monster. The monster was meant to provide three elements: limit to the overall game length, provide a group-oriented motivation factor (beat the monster), and add a small amount of fantasy to the game experience. The monster was excluded from MiBoard in an attempt to simplify gameplay.

Another important difference between the games is that iSTART: The Board Game allows the reader to choose between two strategies while MiBoard specifies a single strategy. Although players are assigned a single strategy in MiBoard, they are provided with the ability to spend their points to change that assigned strategy. This change was implemented as an attempt to diversify students' strategy use, rather than always letting them choose the "lesser of two evils". MiBoard requires students to utilize a variety of strategies and ensures that strategies are not duplicated in consecutive turns for each individual. The developers also hoped this change would encourage players to use their points in a more game-like manner (use them to change strategies, rather than just earn points for no reason).

The last and most noticeable difference between iSTART: The Board Game and MiBoard occurs within the debate portion of the game. Participants in iSTART: The Board Game were allowed to have natural, verbal conversations while participants in MiBoard interacted via chat room, As an attempt to limit off-topic discussion, participants were limited to only three responses per debate session. Also, as an attempt to prevent one participant from delaying the whole game at the chat session, the chat session was limited to three minutes (this time limit approximated the onset of off-topic behavior during user testing). While limiting the time and number of responses was designed to increase the game pace and limit off-topic discussion, these limits may have inadvertently hindered other successful discussion and interaction that was inherently present in the face-to-face version of iSTART: The Board Game.

**Current Study**
The current study investigates user experiences with MiBoard. Specifically, items similar to those used in Rowe (2008) were developed to assess the participants' attitudes and opinions toward MiBoard. These responses were then compared to the results from Rowe (2008). The current study is a modified replication of the iSTART: the Board Game study. 22 participants from an urban University received the same abbreviated iSTART training (lectures only), interacted with MiBoard Game for 30 minutes, and answered questions about their experience with the system.

Contrary to the previous results from iSTART: The Board Game, the participants who played MiBoard did not seem to share the same experiences. Those students who played MiBoard did not enjoy the interaction, found the game to be slow, thought it was somewhat frustrating, and that it did not help them learn (See Table 2 for means).

Table 2. MiBoard Evaluation Responses (ratings from 1 to 6, higher scores indicate agreement)

| Statement | Mean | SD |
| --- | --- | --- |
| The Game Was Fun | 2.14 | 1.39 |
| The Game Improved Strategy Knowledge | 3.77 | 1.60 |
| The Game Was Frustrating to Use | 3.81 | 1.79 |

Additional measures were collected to further assess the gameplay experience. One of the most notable findings was the participants' response to the item. "I would play this game again" (M = 2.09, SD = 1.41). This low score demonstrates the users' lack of enjoyment and disinterest in the game as a whole.

**Discussion**
Results from the two studies were in stark contrast with each other, especially considering that MiBoard was designed as a computerized translation of iSTART: the Board Game. When compared to MiBoard, participants in iSTART: The Board Game had a fun experience, felt the gameplay was easy, and learned useful information. Perhaps the most telling set of responses was the fact that most participants did not want to play the game again (unfortunately this question was not included in the first study).

Possible reasons for these differences stem from the computer-based implementation of iSTART: The Board Game. Specifically, the biggest difference was that participants in iSTART: The Board Game were allowed to conversationally discuss the strategy in a free-flowing verbal conversation until they felt as though they had come to an agreement or an impasse. Participants were not limited in their responses. Instead, the only limit was how long they were willing to discuss the differences. As Rowe (2008) reported, this discussion is where the learning takes place. In MiBoard, however, the discussion had to be computer-based, and therefore took place within a chatroom environment. One problem of the chat system is that it is limited both by the participants' typing skills as well as the system (i.e., limiting number of responses). The simple transition from talking to typing may have hindered the conversational flow during discussion (not to mention the limit on the number of responses). The original reason for limiting the number of turns for discussion was to reduce the amount of off-topic discussion. Unfortunately it may have had a much more detrimental effect by handcuffing the discussion before it even started. It is even possible that off-topic discussion could be important to socially oriented games. Additional problems with the MiBoard chat were noticed: participants had trouble keeping up with the conversation, participants became bored due to the long intervals between responses (while someone was typing), the participants attempted to use the chatroom when it was not available, and sometimes the chat was ignored altogether.

Another contributing difference between iSTART: The Board Game and MiBoard is that the participants did not have a physical presence available for social cues. When playing iSTART: The Board Game, participants can see that a player is reading, thinking, or producing a self-explanation. When the same activities occur in MiBoard, other players see only a static screen. MiBoard attempts to address this issue by displaying messages telling other participants that the reader is busy and that they must wait for them to complete their self-explanation before continuing, but the other participants often do not pay attention to the messages (no matter how obvious they are) and attempt to continue their gameplay. Because there are no available actions during this phase, frustration seems to escalate and the participants' feelings toward the game decline.

**Implications**
These findings are important to the ITS community for multiple reasons. As the ITS field grows rapidly, developers seek proven methods to increase system performance (both in terms of learning as well as engagement). One common methodology for developing an effective tutoring system is to base the design on an effective real-world solution (Jackson, Dempsey, & McNamara, in press). However, this real-world to digital-world translation poses many problems, including those issues addressed here. First, converting a physical system into a computer-based equivalent is not as simple as copying the components into a developer pane. In fact, the strategies that worked in the physical presence program may not even be applicable to a computer-based implementation and completely novel approaches may need to be explored. Second, individual differences play a large role in gameplay. There are likely many performance differences for specific in-game actions (i.e., typing speed, highlighting functions, spatial awareness, etc.) Within MiBoard, additional instructions and a help function (with a pointing hand for each action) were added to ease in-game navigation. Third, in addition to individual ability differences, there are differences in the desire for games in general. Previous research has shown that prior expectations and desires may have more of an impact than individual abilities (Jackson, Graesser, McNamara, 2009)..

In the case of MiBoard it was found that the pace of the game may have negatively contributed to the overall experience. Participants who were waiting for actions to occur often commented that the system was unresponsive and that they wished to move on to some goal-seeking activity, such as fixing the unresponsiveness or even remedying their own need for interaction (i.e., moving on to a different activity).

**Future Directions**

It appears that the MiBoard system must be drastically redesigned if it is to be used with iSTART. Because of the issues presented here, the challenge for the future is to create a practice module for iSTART that incorporates the same effective principles of iSTART extended practice (e.g., computerized environment, quick succession of practice repetitions, no waiting between trials, etc.) within a game setting. Without completely abandoning the collaborative multiplayer aspect of iSTART: The Board Game, we have begun development on a new game (Self-Explanation Showdown) that incorporates the lessons learned from the current study.

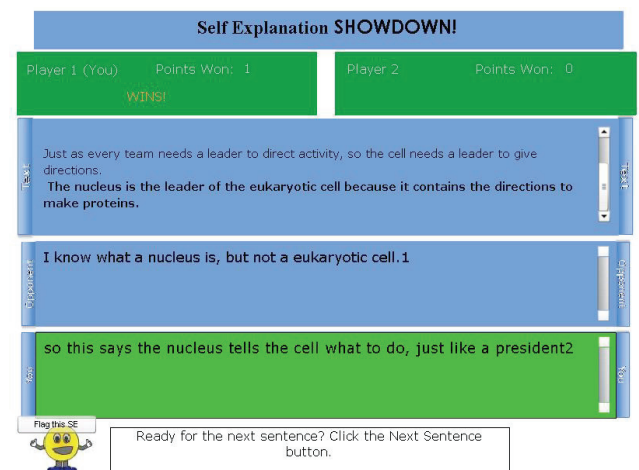

Figure 3. Self Explanation Showdown Interface.

Self-Explanation Showdown is a rapid-fire multi-player game that is intended to increase the engagement through success/failure comparisons against another human opponent. In the game, players both compose a self-explanation based up the same text and same target sentence at the same time (eliminating, or drastically reducing, any wait time). The players' self-explanations are scored using the iSTART algorithm, and then they are directly compared in a showdown. This new game design virtually eliminates the previous problems found in MiBoard, and a new study is soon to be started.

**Conclusions**

The research discussed here will hopefully aid ITS designers as they consider new system designs or improvements to an existing system. The original goal was to create a new form of strategy practice that afforded a long-term and engaging interaction. This goal was achieved through iSTART: The Board Game (Rowe, 2008), but later failed as this same game design was converted into digital format. Ultimately, important lessons of design were learned, and the failures potentially provided more insight than the successes.

**Acknowledgements**

This research was supported by the Institute for Education Sciences (IES R305A080589; IES R305G020018-02). Any opinions, findings, and conclusions or recommendations expressed in this material are those of the authors and do not necessarily reflect the views of the IES. Special thanks to John Meyers for helping in this study.